# A Fast-and-Effective Early-Stage Multi-level Cache Optimization Method Based on Reuse-Distance Analysis

Cheng-Lin Tsai
National Tsing Hua University
Hsinchu, Taiwan
c884523@gmail.com

Ren-Song Tsay
National Tsing-Hua University,
Hsinchu, Taiwan
rstsay@cs.nthu.edu.tw

*Abstract* — In this paper we propose a practical and effective approach allowing designers to optimize multi-level cache size at early system design phase. Our key contribution is to generalize the reuse distance analysis method and develop an effective and practical cache design optimization approach. We adopt a simple scanning search method to locate optimal cache solution in terms of cache size, power consumption or average data access delay. The proposed approach is particularly useful for early-phase system designers and is verified to be 150 to 250 times faster than the traditional simulation-based approach. In addition, we also introduce a simplified analytical model and provide designers insights about how cache design parameters may affect the expected results. As a result, designers can make adequate decision at early system design phase.

*Keywords—multi-level caches; reuse distance;*

## I. Introduction

As CPU performance reaches a plateau, memory subsystem design has become a determining factor for system performance optimization. For memory subsystem optimization, a proper on-chip multi-level cache design is most crucial to reducing the frequency of accessing data from off-chip memories by CPU. An optimal cache design would provide sufficient capacity to store frequently used data, or working data set, while being as small as possible in size for minimal hardware cost and power consumption.

Traditionally, designers perform cache optimization based on simulations of certain targeted applications [1, 2, 3, 4]. Designers evaluate system performance, such as cache hit/miss rate, IPC (Instructions per Cycle) or total energy consumption, of applications on simulated target systems of various cache configurations. Although thorough simulations can provide precise final performance information, generally they are extremely time consuming due to limited simulator speed. Additionally, brute forced simulation approach gives no design insights and laborious work has to be repeated for each design attempt.

Apart from cache size optimization, study on *replacement policy* is also of interest for miss rate improvement. It determines which data in the cache is to be replaced by a new one in case of space conflicts. A simple replacement policy just picks the most infrequently used data to be swapped out. In the literature, many *sophisticated* policies have been proposed [8, 9, 10, 11, 12], mainly for shared last-level cache (LLC) such as that in the Core i7 [23]. However, a general conclusion is that little improvement is accomplished via replacement policy change once the given cache size is large enough [9, 19]. In short, the prime focus for cache design shall be on cache capacity optimization. Specifically, at early design stage, it is of higher priority to find the optimal cache size rather than the replacement policy to achieve fast convergence on system optimization. To meet that end, it is essential to have a fast and effective method to explore optimal cache size for given target applications.

In 1970, Mattson et al. [5] proposed using *reuse distance* histogram as a new way to precisely compute cache hit/miss counts. Given a memory access sequence, each data access is associated with a reuse-distance number, which is defined as the number of accesses of distinct data memory addresses between current data access and the last access to the same data accessing address. Figure 1(a) depicts an example of reuse distance calculation of a memory access sequence. Note that in between the 1$^{st}$ and 6$^{th}$ accesses to the same element *a,* only two unique elements *b* and *c* are accessed. Therefore, the reuse distance of the 6$^{th}$ data access is **2.** The only special case is for the very first access of a data element. Its reuse distance is denoted as infinity, since there is no previous access of the same element. After collecting the occurrence count of each reuse distance number, we construct a reuse distance histogram shown in Figure 1(b), where the horizontal-axis indicates reuse distance number and the vertical-axis is the count of each reuse distance number.

A significant value of the reuse distance histogram is that it can be used to precisely compute the cache hit/miss rate for fully associative, LRU (Least Recently Used) replacement policy cache. According to Mattson et al. [5], the total number of cache hits is simply the sum of the counts of those reuse distances less than the targeted cache size. For instance, if the target cache size is 1KB (i.e. $2^{10}$ bytes) as indicated by the red line in Figure 1(b), then the sum of the counts of reuse distances to the left of the red line is the total hit count. In other words, we can fast and effectively get the cache hit/miss count with no need to perform lengthy simulations. In contrast to long simulation runs, the reuse distance histogram calculation is relatively short. Particularly, lately there are many improvements on accelerating the reuse distance histogram computing based on mathematical methods or parallel programming techniques [6, 7].

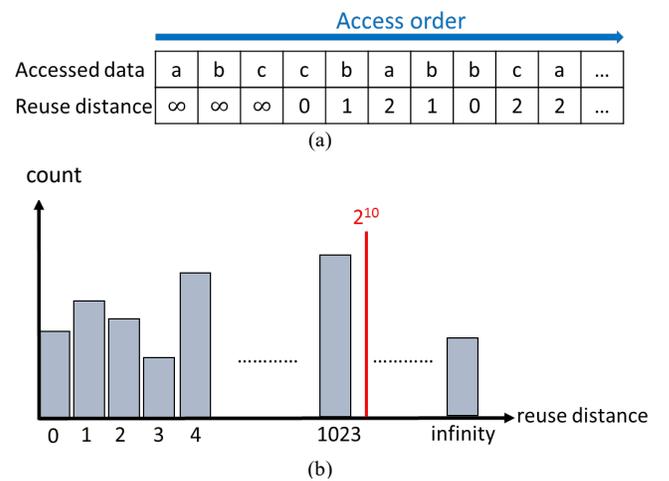

Figure 1: (a) An example of reuse distance calculation of a memory access sequence. (b) An example of reuse distance histogram.

When applying the reuse-distance idea to multi-level cache designs, such as the two-level cache, generally it is suggested to run simulations to collect traces from level-1 (L1) cache misses so that a second level reuse distance histogram can be produced for level-2 (L2) cache evaluation. In this case, the reuse distance histogram must be recomputed whenever the L1 size is changed. The issue is that the Mattson's definition of reuse distance is taken in a one-level view and it requires a generalization in order to be applied to today's mainstream multi-level cache designs. We hence propose in this paper a simple yet effective approach for optimal multi-level cache designs by extending the application of the reuse distance histograms such that there is no need to compute reuse distance at every level but simply the top-level reuse distance histogram is sufficient.

The contributions of this paper are the following: We propose a fast and effective multi cache level's hit/miss count computation and cache size optimization method based on one-level reuse distance histogram. Furthermore, we devise a simple analytical model that system designers can use at early design phase and get the insight on which cache design parameters are critical for optimal results.

The rest of this paper is organized as follows. In Section 2, we first review and discuss related work and then we present in Section 3 a method of using reuse distance histogram for miss rate estimation at each cache level. In Section 4, we propose a simplified model for analytical optimization. In Section 5, we demonstrate the accuracy of our approach by showing experimental results and have a discussion in Section 6 and a brief conclusion in Section 7.

## II. Related Work

Cache performance optimization has always been an important research topic and was intensively studied in the past. In general, there are three cache optimization approaches. The first is to use extensive simulations to explore the best cache designs. The second one is mainly focusing on selecting a good replacement policy and attempts to keep frequently used data in a pre-specified cache architecture. Finally, the reuse distance histogram analysis method was proposed a while back but is getting more attention only recently. Nevertheless, the reuse distance method remained as an analysis tool rather than an optimization tool. We are the first one to apply the reuse distance analysis for multi-level cache optimization and extend the reuse distance approach to be a practical cache design tool. Details of these three approaches are discussed below.

*Design Space Exploration through Simulations*

Typically, at early cache design phase, extensive simulations are conducted to explore best cache designs for target applications. For instance, Jaleel applied different application workloads to characterize memory performance [1]. A few others intensively executed simulations on caches with various design parameters and collected MPKI (Misses per Kilo-Instructions) results to determine cache sensitivity to target applications [2, 3, 4]. In general, the cache design parameters include N-way set associativity, cache line size and replacement policy.

Although the simulation approach provides accurate and detailed results, a critical issue is that the excessive long simulation time is simply impractical for modern day high complexity MPSoC (Multi-Core System-on-Chip) designs mainly due to low simulator speed. Another shortcoming of the simulation approach is that it provides little insight for design guideline as it gives no direct understanding of the optimization factors and causes of critical behaviors. Therefore, arduous simulation process has to be repeated for every new design.

*Replacement Policy Optimization*

Another approach focuses on replacement policy optimization for cache performance. Due to limited cache size, a cache controller needs to provide an intelligent replacement policy for placing new data. A good replacement policy attempts to keep frequently used data in cache and have the least used data replaced.

However, it is observed that in modern multi-core and multi-level cache systems, the basic replacement policies, such as FIFO, RANDOM, MRU, and LRU, may perform poorly on LLC, due to the fact that LLC is usually shared by different applications running on different cores. In case of capacity deficiency, the sharing of LLC often leads to capacity contention or thrashing effects. Additionally, the lower level caches often have already absorbed the temporal locality and hence lead to the fact that the data in LLC reflects little locality. Therefore, the policies work for lower level caches often perform poorly at higher level caches.

Prior studies [8, 9, 10, 11, 12] confirmed that memory access patterns are different in LLC as a result of cache sharing of mixed applications. To address this issue, some proposed heuristic replacement policies for adequate victim cache replacement based on simulation analysis of target applications' data access behaviors [8, 9]. Some attempted cache partitioning strategy in order to balance cache space usage for each core [10, 11, 12]. Nevertheless, these proposed heuristics are all based on the commonly used LRU replacement policy as a base line for improvement.

Generally, replacement policies all require extensive experiments to know how cache size change may affect the performance. In fact, Wu et al. conclude that mainly cache size determines the final performance and replacement policy serves only to fine tune performance [9, 19]. Moreover, LRU is still the most adopted policy in all levels of modern multi-level cache designs. Therefore, for early system-level design phase, LRU can effectively be used as the reference policy for evaluating the best cache size for a target application.

*Reuse Distance Analysis*

Mattson et al. [5] in 1970 proposed a fast and accurate reuse-distance-based model for hit/miss count estimation for fully-associative LRU cache of any size with no need of running simulations. However, the reuse distance histogram computation can still be time-consuming, although it is much faster than running exhaustive simulations of all possibilities. To effectively shorten reuse distance computation time, *StatCache* [6] proposed to compute only on sampled partial memory access sequence with a sampling rate as low as $10^{-4}$ and adopted a probabilistic model to evaluate hit/miss counts accurately from the run-time statistics.

Furthermore, *PARDA* [7] leveraged the computing power of multi-core systems and proposed a parallel reuse-distance computing technique based on a tree-based data structure and reduced the time complexity from $O(M)$ to $O(logM)$, where $M$ is the number of different memory references.

Note that all these reuse distance analysis approaches were limited to one-level cache analysis, while multi-level cache is the norm for today's designs. In this paper, we extend the reuse distance analysis approach to be applicable to multi-level cache designs and thereafter develop a systematic method to determine the best cache architecture configuration. Our proposed analytical model also provides optimization insight to facilitate system designers for quick and effective early design decisions.

Despite the fact that the reuse distance histogram is derived from fully associative cache, our experiments show that the practical 8-way or 16-way set caches give almost the same miss rate results. In addition, we also verify true that Pseudo-LRU, a low-overhead version of LRU and the most widely adopted replacement policy in commercial processors [18], produces little difference to the LRU algorithm. Therefore, our proposed approach is proven effective for practical early system cache designs.

### III. The Reuse-Distance Approach for Multi-level Cache Designs

In general, there are two types of multi-level cache architectures: exclusive and inclusive [13]. We first discuss the difference of their data placement behaviors, and then how that reflects on the reuse distance histogram. We then propose a method using the reuse distance histogram for hit/miss count estimation of caches of any size in any level for both exclusive and inclusive type of caches. The estimation is verified to have negligible error rate by the experiments in Section 5. First, we explain the difference of typical exclusive and inclusive caches and then elaborate how to apply our proposed reuse-distance approach.

*Exclusive/Inclusive Cache*

We use Figure 2 to illustrate the difference of exclusive and inclusive caches. For exclusive cache architecture as shown in Figure 2(a), the higher level cache does not replicate data of lower level cache. When a data request is issued by CPU, cache system always searches L1 first for data availability. If L1 cache hits, or the data is available, the system directly returns the data; otherwise, it goes on searching the higher level cache, i.e. L2 in this case. If L2 misses, or the data is not in L2, then it will fetch the data from next level cache or memory to fill in L1. On the other hand, if L2 hits, or the data is in L2, then the data is moved to L1 from L2. For either case when L1 misses, a victim data is picked from L1 and pushed back to L2 in order to make room for the new data if L1 is fully occupied.

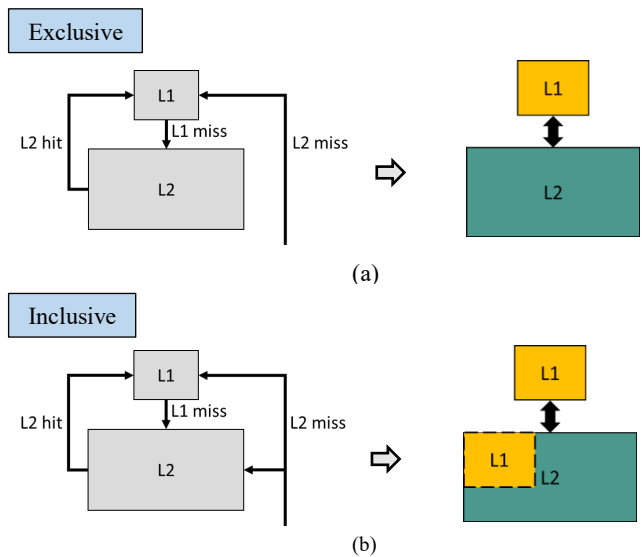

Figure 2. The data access path and placement of (a) an exclusive cache and (b) an inclusive cache.

The inclusive cache is illustrated in Figure 2(b), for which the higher level cache always has a data replica of the lower level cache. For L1-hit case, the data requested is accessed directly from L1, same as that in the exclusive cache. For the case L1-miss but L2-hit, the data is replicated from L2 to L1. When L2 also misses, the data is fetched from next level cache or memory and copied to both L1 and L2. Essentially, L2 always has a copy of L1 content.

In general, the exclusive cache architecture has better capacity utilization but the inclusive cache has less coherence overhead when applied in the multi-core systems.

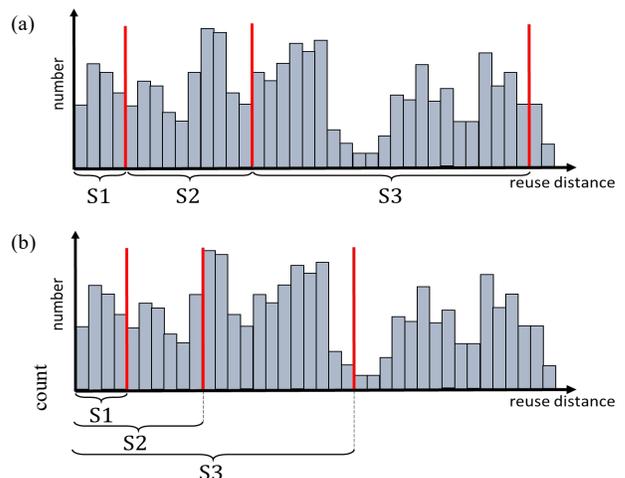

Figure 3. Compute miss counts for multi-level caches using reuse distance histogram, (a) for **exclusive** caches (b) and for **inclusive** caches, where $S_i$ is the size of the *i*-th level cache.

*Applying Reuse Distance in Multi-level Designs*

We now discuss how to extend the traditional reuse-distance analysis method to multi-level cache designs.

For the following discussions, we assume that a specific data access *a* is associated with a reuse distance value $R_a$. According to the reuse distance definition, there are $R_a$ counts of unique data requests to the cache in between the current and previous accesses of the data *a*. For convenience, we let $S_i$ be the size of *i*-th level cache $L_i$.

We first discuss the case of multi-level exclusive cache designs. For this case, a victim data is pushed to the higher level cache. Therefore, if $R_a$ is greater than $S_1$, the current access cannot find *a* in L1 and it causes a cache miss. In general, if $R_a > (\sum_{j=1}^{i} S_j)$ then the data *a* is missing in L1,…, Li. Therefore, we may conclude that a miss in Li occurs if the reuse distance is greater than $(\sum_{j=1}^{i} S_j)$, and hence we can accumulate the counts of those reuse distances to the right of $(\sum_{j=1}^{i} S_j)$ to get the total count of Li misses as illustrated in 3(a).

In contrast, for the multi-level inclusive cache, the content of the lower level cache always takes space in the higher level cache. Thus, a data miss in Li occurs if its reuse distance is greater than $S_i$. Therefore, we accumulate the counts of those reuse distances to the right of $S_i$ to get the total count of Li misses as shown in Figure 3(b).

According to the above discussions, we find that we may use the reuse distance histogram for evaluating hits/misses of multi-level cache system. We actually verify that the so resulted estimation is very accurate as verified by experiments. In the following, we further extend the findings and develop a systematic cache optimization method particularly adequate for early system design phase.

## IV. A Systematic Cache Optimization Method

*Scanning Search for Optimal Designs*

Equipped with the accurate hit/miss count estimation method based on the reuse distance analysis, we now discuss how to find the optimal multi-level cache designs using a scanning search method.

Once the microarchitecture is determined, the CPU computation performance is pretty much fixed and what left for system performance optimization is the data access delay through cache hierarchy to main memory. If the cost is not a concern, an obvious solution for the best system performance is to have the fastest memory (or cache) fully deployed. Nonetheless, the cost of such a system is simply impractical.

Therefore, a practical cache design objective is either to minimize the total cache implementation cost *c* while meeting a target average data access delay value *T*, or to minimize the average data access delay *t* while having the total cache cost constrained by a target value *C*.

We first discuss the case of minimizing cache cost on one-level cache architecture and later generalize the solution to cover multi-level cases. Suppose that the size of the L1 cache is $x_1$, the cache cost function is $c(x_1)$, cache miss rate is $M(x_1)$ and target average data access delay value is *T*. Assume that the average execution time per instruction under ideal cache is $CPI_{base}$ and the miss delay penalty is *m*, then the cache optimization problem can be formulated as the following:

$$min: \quad c(x_1)$$
$$s.t. \quad t = CPI_{base} + m * M(x_1) \leq T$$

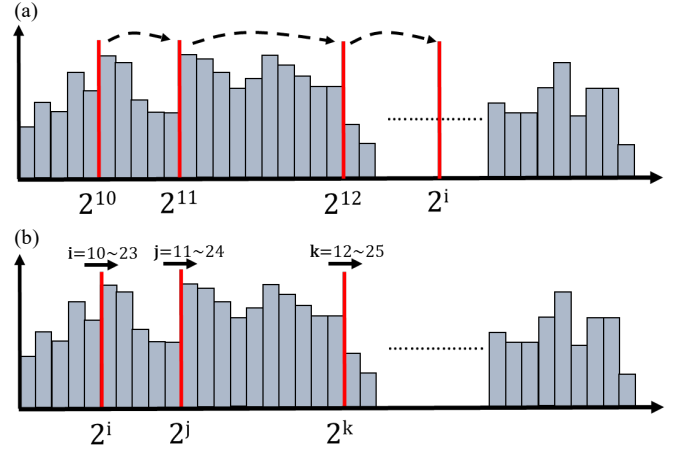

Figure 4. An illustration of the scanning search method. (a) Scan through the possible size choices and evaluate the corresponding cost for optimal one-level cache design; (b) Scanning test on three-level cache

Note that $c(x_1)$ shall be a monotonically increasing function and $M(x_1)$ a monotonically decreasing function. Due to the monotonicity of both the cost and miss rate functions, the solution to this problem is in fact quite trivial. We essentially take the smallest cache that can meet the access delay constraint, i.e. $CPI_{base} + m * M(x_1) \leq T$. In other words, we have $x_1 \geq M^{-1}(\frac{T-CPI_{base}}{m})$.

Since the inverse function $M^{-1}()$ is not trivial, now we discuss how to find the optimal cache size in practice. In general, the miss rate function is not linear due to the fact that the reuse distance function is not linear. Therefore, its inverse function $M^{-1}()$ is not trivial. Nevertheless, we observe a fact that the cache size is always of the power of *2*, i.e. $2^i$, in unit of byte or word. Additionally, since the value $2^i$ grows exponentially along with *i*, practically we have limited number of cache size choices. Consequently, an easy solution to find the optimal cache size is simply to perform a scanning search through the few possible choices, such as *i*=10 to 23 as shown in Figure 4(a). The smallest *i* that satisfies the average data access delay constraint shall be the optimum cache size solution.

We now extend the formulation to handle *n*-level cache architecture optimization as the following:

$$min: \quad c(x_1, \dots, x_n)$$
$$s.t. \quad t = CPI_{base} + \sum_{i=1}^{n} m_i * M_i(x_i) \leq T$$

where $m_i$ is the miss delay penalty and $M_i(x_i)$ is the average miss rate function at level *i*. Although the cost function $c()$ and the miss rate function $M_i(x_i)$ all preserve the monotonicity property, the solution is not as straight forward as that of the one level cache case. However, as discussed previously the choices of cache at each level is essentially of a limited number as the size has to be of a value of the power of 2. Therefore, Figure 4(b) shows a simple scanning search method which basically evaluates all possible combinations in the multi-level case.

Specifically, in our experiment we evaluate the three-level designs of size combinations $(2^{10}\sim2^{23})*(2^{11}\sim2^{24})*(2^{12}\sim2^{25})$, and the total number of possible cache combinations is merely 560.

If the total cache power consumption is the objective function as modeled in [22, 23], then we have the $n$-level cache optimization formula

$$\min: \quad P = \sum_{i=1}^{n} [p_{static}^i(x_i) + p_{dynamic}^i(M_{i-1}(x_{i-1}))]$$

$$s.t. \quad t = CPI_{base} + \sum_{i=1}^{n} m_i * M_i(x_i) \leq T$$

Note that the total power $P$ consists of two parts. The first one is the static power $p_{static}$ which dissipates energy continuously due to leakage current and is roughly proportional to the cache or memory size. The other one is the *dynamic power* part which is contributed by logic switching activities caused by data access operations. Therefore, $p_{dynamic}$ is roughly proportional to the miss count from the last lower level cache, while the dynamic power consumption of the first level cache is proportional to the data accesses from CPU. Obviously, the misses from the last level cache will affect the dynamic power consumption of the main memory. In practice, the power consumption model can be generated from tools such as CACTI [22] or from detailed circuit simulations.

Again, once the optimization formulation and the cost function are determined, we can easily use the scanning search approach to locate the optimal cache designs based on the reuse distance histogram as discussed previously.

Similarly, if the objective is to minimize the average data access delay subject to a cost constraint, then we can formulate the problem as the following.

$$\min: \quad t = CPI_{base} + \sum_{i=1}^{n} m_i * M_i(x_i)$$

$$s.t. \quad c(x_1, \dots, x_n) \leq C$$

We then can again apply the scanning search to find the best design.

*Analytical Model based-on Reuse Distance*

Now we assume a simplified reuse distance histogram in the form of a step function. Although the assumption is idealistic, the derived analytical optimum solution provides a good design insight, which is extremely helpful for system designers at the early design phase.

We first assume that the reused distance histogram function is a step function with a non-zero constant count number only in the region from reuse distance zero to a maximum reuse distance cut off value $D$ as shown in Figure 5(a).

Then for the one-level cache case, the miss rate $M(x_1)$ simply equals to $(D - x_1)/D$. If we assume that the size cost function $c(x_1) = ax_1^2$, a quadratic function and $a$ is a positive constant. In other word, we have

$$\min: \quad c(x_1) = ax_1^2$$

$$s.t. \quad CPI_{base} + m * (D - x_1)/D \leq T$$

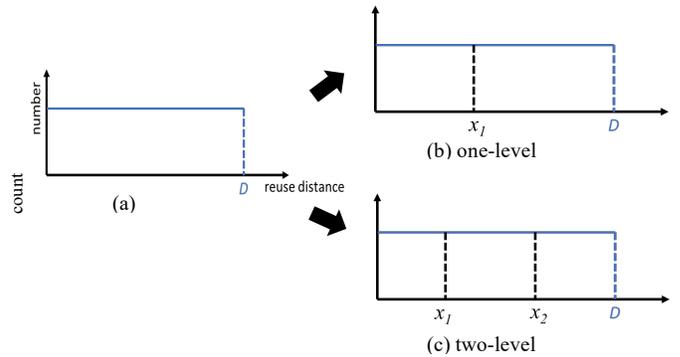

Figure 5. A simplified analytical model of reuse distance histogram.

The above quadratic formulation has an analytical optimum solution, i.e.

$$x_1 = D\left(1 - \frac{T - CPI_{base}}{m}\right)$$

This analytical solution tells us that if the target data access delay $T$ is relaxed to be a larger value, then the cache can be chosen to be smaller. In case that $T$ is really large, or data access performance is not an issue, then essentially cache is not needed, as $x_1$ shall be equal to zero.

On the other hand, if the instruction computing time $CPI_{base}$ or miss rate $m$ deteriorates to be a larger value, then the cache has to be enlarged to keep the data access delay conform to specification, according to the above equation.

Interestingly, if the instruction computation $CPI_{base}$ is fast (i.e. small) enough or the miss penalty $m$ is small enough, we may not even need cache at all. This is the case if the L1 access latency is too long or the DRAM access latency is very fast, then L1 is unable to help system performance improvement and is simply a waste.

Now for the two-level inclusive cache case as shown in Figure 5(b), we have

$$\min: \quad c(x_1, x_2) = a_1 x_1^2 + a_2 x_2^2$$

$$s.t. \quad CPI_{base} + m_1(D - x_1)/D + m_2(D - x_2)/D \leq T$$

Note that the unit cost $a_1$ of level-1 cache is generally larger than that of the level-2, i.e. $a_2$, in practice. Then solving the above formulation, we have the optimum solutions

$$\begin{cases} x_1 = a_2 m_1 P \\ x_2 = a_1 m_2 P \end{cases}, \text{ where } P = \frac{D(CPI_{base} + m_1 + m_2 - T)}{a_1 m_2^2 + a_2 m_1^2}.$$

With the above analytical solutions, we may derive a few interesting design insights. For example, if the unit cost $a_2$ of L2 increases, then L2's size has to become smaller and L1 larger in order to balance the delay penalty conflict. Conversely, if the unit cost $a_1$ of L1 increases, we should have a smaller L1 and a larger L2. On the other hand, if the miss penalty $m_2$ of L2 increases, L2 should increase in size to reduce access delay while L1 has to be smaller to abridge cost increase.

The most interesting case is when the miss penalty $m_1$ of L1 increases, both L1 and L2 have to increase in size rather than just L1 increases, mainly due to the fact that $a_1 > a_2$. In this case, the total cost is not minimized if only L1 takes responsibility for reducing delay increase. Hence, only larger caches of both levels can achieve minimal result.

Although, the uniform step reuse distance distribution assumption is somewhat simplistic, the main purpose is to obtain design insights through the analytical solutions and understand how each design parameter may affect the final design cost. In reality, scanning search approach shall provide more realistic solutions.

## V. Experiments

*Methodology*

To verify our proposed optimization method, we adopt SPEC CPU2006 [17] benchmark suite as the target applications, adapt the cache model of SimpleScalar [16] to support three-level cache architecture, and evaluate both inclusive and exclusive models. All experiments are performed on a host machine of *Intel Xeon-E5-2650 2.00GHz.*

For the experiments, we first use SimPoint [14] to find a representative execution region in each benchmark and then apply Pin [15], a dynamic instrumentation tool, to collect traces of one billion memory accesses of the representative region. Then, we apply the traces to evaluate hit/miss rate of different multi-level cache architectures of various cache size combinations.

For fair comparison, we first execute the trace-driven simulation on SimpleScalar and produce golden results for comparison. To fully utilized the multi-core host machine, we adopt the parallel reuse distance histogram computing method [7]. Then we apply our proposed reuse distance estimation method discussed in Section 3 for miss rate estimation. Finally, we compare our estimated results with the golden simulation results and verify the accuracy of the proposed approach.

TABLE I. Size range for each cache level

| | for both Inclusive/Exclusive | |
|---|---|---|
| L1 | $2^i$ bytes, $10 \leq i \leq 23$ | $i, j, k \in \mathbb{N}$, $i < j < k$ |
| L2 | $2^j$ bytes, $11 \leq j \leq 24$ | |
| L3 | $2^k$ bytes, $12 \leq k \leq 25$ | |

Regarding the cache size configurations, we list in Table I the cache size range of each level. Thus, we test each case on a total of 560 possible combinations. Note that the LRU policy is adopted for all tests and the default cache line size of each level is 64 bytes.

*Evaluation Results*

For exclusive cache architecture, our reuse-distance-based method basically produces the same results as those from the golden simulations. The main reason is that victim data pushed into the next level always resets its LRU value to 0 and hence the *relative* LRU value is equivalent to that as if we treat both L1 and L2 as one cache. Therefore, the reuse distance histogram can provide 100% accurate hit/miss rate evaluations for multi-level exclusive cache designs.

In contrast, for inclusive cache designs, the victim data is replaced without affecting the replica in its next level cache and hence its LRU value is outdated. As a result, it causes slight inaccuracy of the LRU values in higher levels. Nevertheless, as shown in Figure 6, the average miss error rate of L2 and L3 is less than 4% and the max error is less than 4.78%.

Since the reuse-distance estimation method assumes ideal LRU, we also perform another test which modifies the inclusive cache implementation simply by inheriting the LRU value of the victim data to the next level cache. Then the reuse-distance-based miss count estimations match perfectly with the golden simulation results.

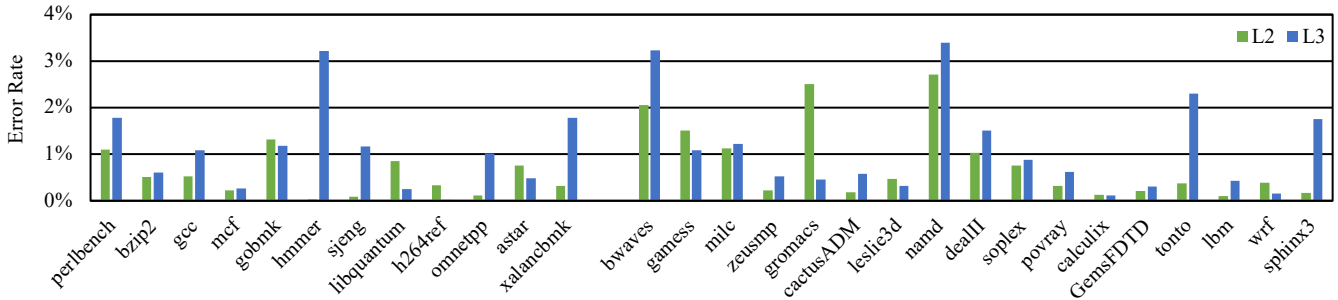

Figure 6. The average error rate of each SPEC benchamek case on a total 560 cache configurations.

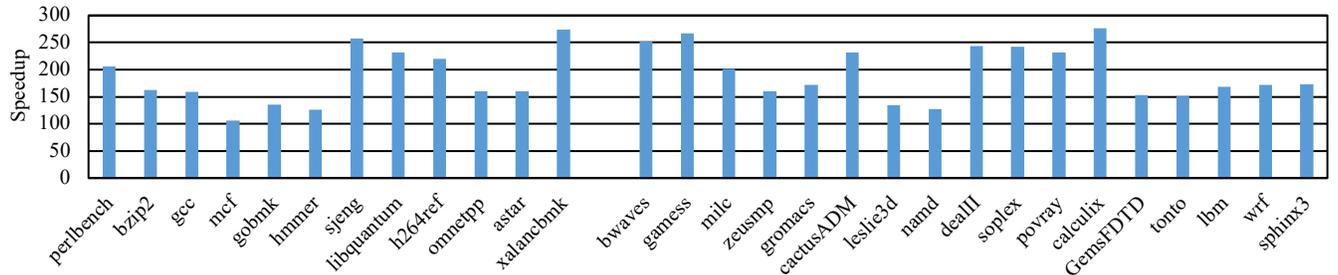

Figure 7. Speedup of our proposed method compared to simlation-based method.

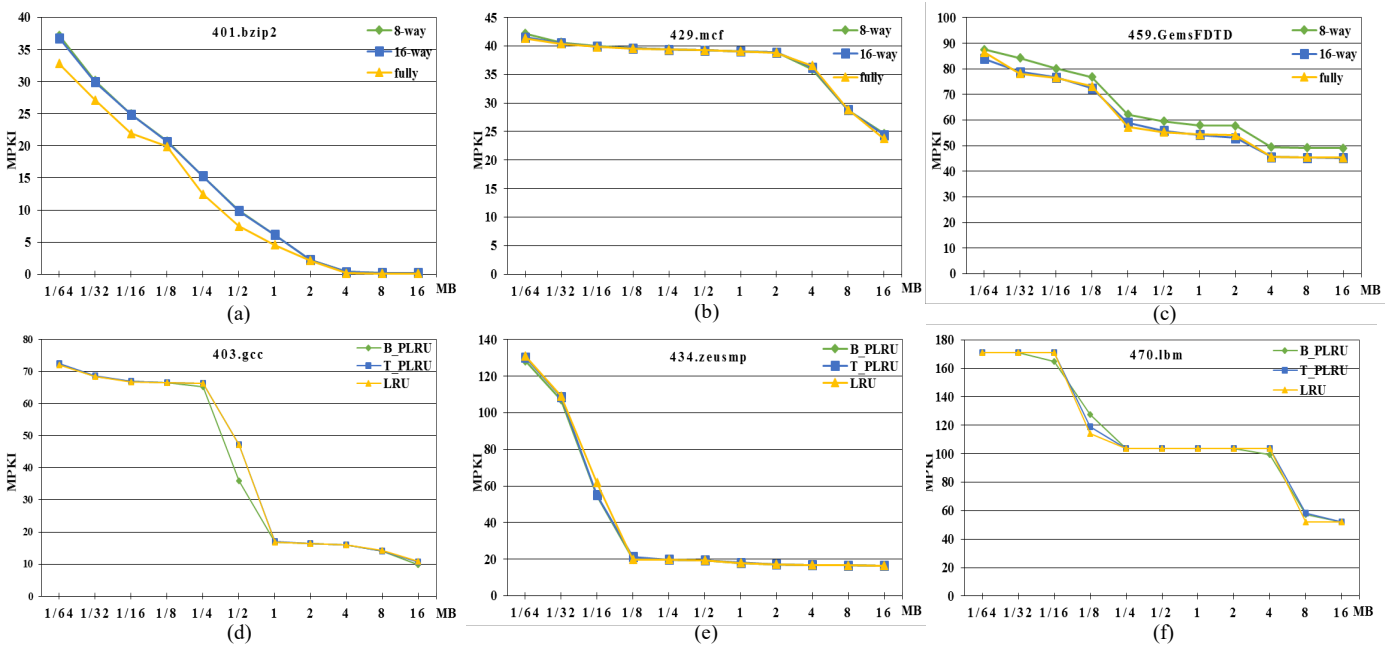

Figure 8. (a)(b)(c) Comparing N-way with fully-associative designs. (d)(e)(f) Comparing PLRU with LRU replacement policies in 16-way cache.

In short, the experimental results support that our proposed method is very accurate for evaluating multi-level cache hit/miss counts. As a result, designers may effectively use the proposed hit/miss rate estimation results for optimal cache designs.

Moreover, we show in Figure 7 the speedup of the total computation time required for 560 simulations and that for reuse distance histogram calculation. Basically the simulation-based approach takes 150 to 250 times longer computation time than our proposed method. In average, it requires in average only 14 minutes to generate a reuse distance histogram and estimate the miss rates. Therefore, the reuse distance approach is practical for cache architecture optimization for early system designs.

*Verify Insensitivity to Replacement Policy and Way-associativity*

Since in practice and particularly in recent years, Intel and AMD ubiquitously adopt 8-way or 16-way cache designs. Additionally, Pseudo-LRU, either bit-based or tree-based, replacement policy is the most commonly used. Therefore, we also conduct experiments to verify whether our proposed reuse distance approach is also good for N-way set associative and Pseudo-LRU cache designs.

For verification, we use MPKI (Misses per Kilo Instructions) for comparison and perform two different tests on each benchmark case. For the first test, we fix the LRU replacement policy but compare results of 8-way set, 16-way set and fully associative configurations. For the second test, we compare the results of bit-based PLRU, tree-based PLRU and the ideal LRU replacement policies in 8-way, 16-way and fully associative caches. Note that for this test we show only the results of 16-way's cases because the others exhibit similar results.

As shown in Figure 8, we observe only little disparity among the results of different way-associativity numbers and replacement policies. Therefore, the proposed reuse-distance-based cache optimization approach is verified to be effective in practice.

*Disucssions*

Our approach can be extended to determine optimal cache block size. Due to spatial locality, increasing block size may decrease miss rate. Therefore, cache designs generally should consider a proper choice of block size. However, if we are to apply our approach we find a fact that the choice of different block size leads to different reuse-distance histogram. For instance, the maximum reuse distance value is halved for doubled block size because the total block number is halved as illustrated in Figure 9(a).

Nevertheless, we may normalize the horizontal axis (i.e. the reuse distance) of the histograms of different block sizes to be in the same 1-byte unit as shown in 9(b). The purpose of aligning the reuse-distance histograms of different block sizes is that we then can easily apply concurrently the scanning search on all histograms and identify the optimal block size.

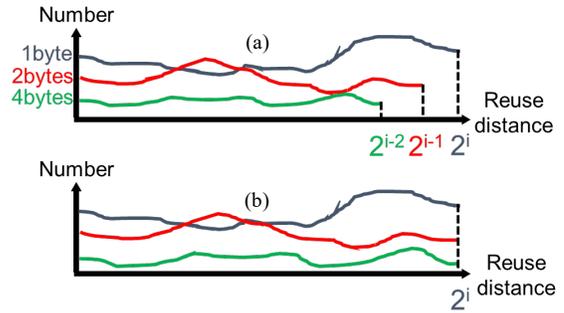

Figure 9: (a) Histograms with different block sizes. (b) Normalized histograms of different block sizes.

Another issue to be discussed is that in this paper we assume both instructions and data streams share the same L1 cache. Practically, there are separated L1 instruction cache and data cache, and shared L2 and L3 caches. For future work, we plan to study the optimization of separated L1 cache sizes and more importantly the general cache optimization of multiple applications contending caches in multi-core machines.

## VI. Conclusion

In this paper, we have proposed a systematic multi-level cache analysis method that can accurately estimate the cache miss rates of all cache levels based on reused-distance histogram. The experimental results show that the multi-level miss rate estimation is only 4% from the golden references in average. With the proposed scanning search approach, designers can efficiently and precisely determine optimal multi-level cache designs. Additionally, we also use a simplified analytical model and derive useful design insights for early system designers. For future work, we plan to further extend the idea for optimal multi-core shared cache designs.